\documentclass{aa}

\usepackage{graphics}

\begin{document}

\def\deg{\ifmmode^\circ\else\hbox{$^\circ$}\fi}                                     
\def\1{\'\i}                    
\def\3{\c c}                   
\def\4{\"u\^e}                           
\def\ao{\~ao\ }                             
\def\oes{\~oes\ }                                                        
\def\th{\thinspace}                   
\def\hb{\hfil\break}                   
\def\cao{\c c\~ao\ }                   
\def\coes{\c c\~oes\ }                               
\def\ct{\centerline}                                    
\def\el{{\it et al.}}                   
\def\bu{$\bullet$}
\def\dy{\displaystyle}
\def\sy{\scriptscriptstyle}                                    
\def\lsim{\,\raise 0.25ex\hbox{$<$}\hskip -0.7em \raise -0.45ex                    
\hbox{$\scriptstyle\sim$}\hskip 0.2em\,}                   
\def\gsim{\,\raise 0.25ex\hbox{$>$}\hskip -0.7em \raise -0.45ex                  
\hbox{$\scriptstyle\sim$}\hskip 0.2em\,}              
\def\llsim{\raise 0.26ex\hbox{$\scriptscriptstyle <$}\hskip -0.5em \raise -         
0.45ex\hbox{$\scriptscriptstyle\sim$}\hskip 0.2em}                   
\def\ggsim{\raise 0.26ex\hbox{$\scriptscriptstyle >$}\hskip -0.5em \raise -         
0.45ex\hbox{$\scriptscriptstyle\sim$}\hskip 0.2em}                 
\def\asolm{\lower 1.1em\hbox{\hskip -1.5em$\sy \Omega_M$\hskip 1ex}}               
\def\asol{\lower 1.1em\hbox{\hskip -1.5em$\sy \Delta\Omega$\hskip 1ex}}                  
\def\a4pi{\lower 1.1em\hbox{\hskip -1.4em$\sy 4\pi$}\hskip 1ex} 
\def\afence{\lower 1.2em\hbox{\hskip -1.7em$\sy \Omega_f$\hskip 1ex}}                                     
\newbox\strutbox                   
\setbox\strutbox=\hbox{\vrule height11pt depth3.5pt width0pt}                   
\def\ordo{\deg\hskip -0.37em -\ }                                    
\def\sep{\to\>\,\,}                   
\def\mag{,\hskip -0.37em \raise 1ex\hbox{$\scriptstyle m$}}                   
\def\beq{\begin{equation}}                  
\def\eeq{\end{equation}}                  
\def\beqy{\begin{eqnarray}}                  
\def\eeqy{\end{eqnarray}}                  
\def\ifi{\Longleftrightarrow}                 
\def\beqsp{\vspace{7.227pt}}              
\def\eeqsp{\vspace{-10pt}\noindent}          
\def\gem{GEM}                     
\def\pin#1{\put(10,4){\circle{12}}\raise 0.2ex\hbox{\hskip        
1.45ex{\small{#1}}\hskip 1.45ex}}                       
\def\pn#1{\put(10,4){\circle{14}}\raise 0.2ex\hbox{\hskip        
0.82ex{\small{#1}}\hskip 1ex}}       
\def\fa{408\thinspace{MHz}}         
\def\fb{1465\thinspace{MHz}}  
\def\fc{$2.3$\th{GHz}} 
\def\fd{5\th{GHz}}      
\def\fe{10\th{GHz}}
\def\ord#1{\raise 0.7ex\hbox{\small\underbar{#1}}}        
\def\ordd#1{\raise 0.8ex\hbox{\tiny\underbar{#1}}} 

   \thesaurus{22     
              (03.13.1;  
               03.13.5;  
               03.13.7;  
               03.19.1;  
               13.18.3)  
               } %
   \title{Spillover and diffraction sidelobe contamination\\ 
in a double-shielded experiment for mapping\\ 
Galactic synchrotron emission}


   \author{C. Tello\inst{1} \and T. Villela\inst{1} \and G. F. Smoot\inst{2} 
		\and M. Bersanelli\inst{3} \and N. Figueiredo\inst{4}
		\and G. De Amici\inst{5} \and M. Bensadoun\inst{6} 
		\and C. A. Wuensche\inst{1} \and S. Torres\inst{7} 
          }

   \offprints{tello@das2.inpe.br}

   \institute{Divis\ao de Astrof\1sica, 
		Instituto Nacional de Pesquisas Espaciais (INPE), CP 515,
			S\ao Jos\'e dos Campos, 12201-970, Brazil
	\and
		Lawrence Berkeley National Laboratory, 
		University of California at Berkeley, 1 Cyclotron Road, 
		Bldg. 50, MS 205, Berkeley, CA 94720, USA
	\and
		Dipartimento di Fisica, Universit\'a di Milano, Via Bassini 15,
		20133 Milano, Italy
   	\and
		Escola Federal de Engenharia de Itajub\'a, 
		Av. BPS 1303, Itajub\'a, 37500-000, Brazil
	\and
		TRW R1/2128, 1 Space Park Drive, Redondo Beach, CA 90278, USA
	\and
		Newfield Wireless, Inc., 2907 Claremont Ave, Suite 111,
		Berkeley, CA 94705, USA
	\and
		Centro Internacional de F\1sica, Bogot\'a, Colombia
             }

\date{Received January 20; accepted June 7, 2000}

\titlerunning{Spillover and diffraction sidelobe contamination}
\authorrunning{C.~Tello et al.}

   \maketitle

   \begin{abstract}
We have analyzed observations from a radioastronomical experiment to survey 
the sky at decimetric wavelengths along with feed pattern measurements in order 
to account for the level of ground contamination entering the sidelobes. 
A major asset of the experiment is the use of a wire mesh fence around the 
rim-halo shielded antenna with the purpose of levelling out and reducing this 
source of stray radiation for zenith-centered 1-rpm circular scans. 
We investigate the shielding performance of the experiment by means of a 
geometric diffraction model in order to predict the level of the spillover and 
diffraction sidelobes in the direction of the ground. 
Using \fa\ and \fb\ feed measurements, the model shows how a weakly-diffracting 
and unshielded antenna configuration becomes strongly-diffracting and 
double-shielded as far-field diffraction effects give way to near-field ones. 
Due to the asymmetric response of the feeds, the orientation of their radiation 
fields with respect to the secondary must be known a priori before comparing model predictions with 
observational data. 
By adjusting the attenuation coefficient of the wire mesh the model is able to 
reproduce the amount of differential ground pick-up observed during test 
measurements at \fb. 

      \keywords{Methods: analytical -- Methods: observational -- 
		Methods: laboratory -- Site testing -- Radio continuum: ISM
               }
   \end{abstract}


\section{Introduction}
The Galactic Emission Mapping (GEM) project (De Amici et al.~\cite{Ami}; 
Torres et al.~\cite{Tor}; Smoot \cite{Sm2}) is an on-going international 
collaboration, presently mapping the radio sky at decimetric wavelengths in order 
to provide a precise understanding of the spatial and spectral distribution of the 
synchrotron component of Galactic emission. In today's cosmological scenario 
Galactic foreground contamination plays a central role. Despite the unprecedented 
success that microwave astronomy achieved in the last decade (e.g.~Smoot et 
al.~\cite{Sm1}; Gundersen et al.~\cite{Gun}; Lim et al.~\cite{Lim}; Davies et 
al.~\cite{Dav}), an unambiguous identification of the level of contamination from 
our own Galactic environment still awaits a more reliable treatment in the face of 
existing data (Lawson et al.~\cite{Law}; Banday \& Wolfendale \cite{Ban}; 
Bennett et al.~\cite{Be1,Be2}; Kogut et al.~\cite{Ko1,Ko2}; 
Platania et al.~\cite{Pla}; Jones \cite{Jon}; L\'opez-Corredoira \cite{Lop}). 

One often-neglected source of contamimation affecting the baseline determination 
of present-day surveys of the radio-continuum of the sky in decimeter 
wavelengths (Haslam et al.~\cite{Ha1, Ha2, Ha3}; Berkhuijsen \cite{Bhj}; 
Reich \cite{Rei}; Reich \& Reich \cite{RR}) is the component of stray radiation 
emitted by the ground when coupled to the observational technique. These 
surveys were obtained with some of the largest single-dish radiotelescopes in the world as 
they scanned the sky over limited angular ranges either along the meridian circle 
or at constant elevation. In order to completely sample the accesible portions 
of the sky, however, low scanning speeds 
(3$\deg$--10$\deg$\thinspace{min}$^{-1}$) were required by the medium 
resolution of these large radio dishes. This requirement introduces striping in the 
maps as a result of 1/f noise enhancement along the scanning direction 
(Davies et al.~\cite{DWG}; Dellabrouille \cite{Del}; Maino et al.~\cite{Mai}). In 
addition, scanning in azimuth can likewise produce horizontal (parallel to right 
ascension) striping due to an horizon dependent ground pick-up through the 
antenna sidelobes. In the GEM experiment we scan the sky from different sites at 
the constant elevation of $60\deg$ with a portable $5.5$-m dish rotating at 
1 rpm. Thus a crucial element of our experiment is the reduction and proper 
accounting of the antenna sidelobe contamination by ground emission. Even 
though we make an effort to minimize and level out the ground emission signal 
by using fixed and co-rotating ground shields (see Fig.~\ref{Fig1}), the sensitivity goal for 
our low resolution sky measurements (S/N$\sim$10) demands  a more 
comprehensive treatment of the role played by diffraction and spillover sidelobes.
The importance of stray radiation corrections in survey experiments has already 
been made clear in the past as, for instance, in Hartmann et al.~(\cite{Har} and
references therein) when applied to the Leiden/Dwingeloo survey of HI in the 
Galaxy (Hartmann \& Burton \cite{HB}). 

\begin{figure}[hbt]
\resizebox{8.8cm}{!}{\includegraphics{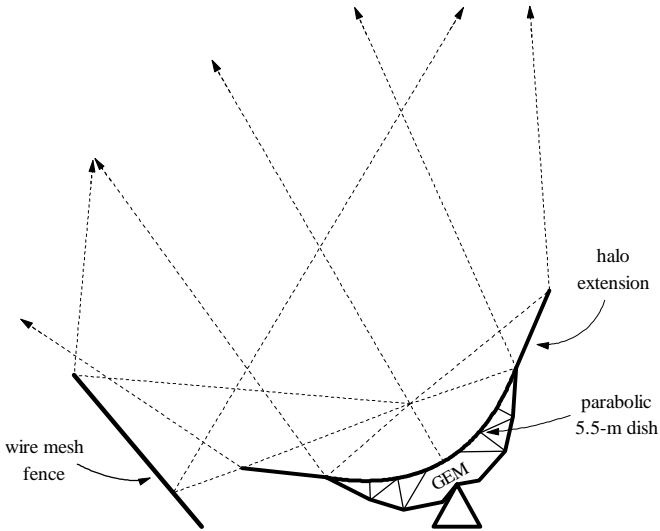}} 
\caption{Schematic representation of a ray-tracing diagram (dotted lines) for the 
double-shielded portable radiotelescope of the GEM project.
	}
\label{Fig1}
\end{figure}

In this article we first demonstrate the effective use of a fixed ground shield 
in levelling out the contamination from the ground (Sect.~\ref{ADGL}) for GEM 
observations at \fb. We then set out to determine the extent of this 
contamination by comparing model predictions of the spillover and diffraction 
sidelobes that overlook the ground behind the shields (Tello et al.~\cite{Te2}, 
from now on Paper I) with differential measurements of the antenna temperature 
toward selected regions of the sky. In order to do so we will rely upon a complete 
radiometric description of the feed (Sect.~\ref{FPM}) and a detailed study of its expected 
performance under different shielding configurations (Sect.~\ref{AGC}). Then we will use 
the near sidelobe pattern (out to some $30\deg$ from axis) of the radiotelescope 
to pin down the proper orientation of the feed pattern with respect to the optical 
axis of the secondary before finally subtracting the differential contributions of the 
atmosphere and the Galaxy (Sect.~\ref{TM}). The latter will be obtained from a template 
sky based on a preliminary GEM survey at 1465 MHz in the Southern sky. A 
summary of the article and its main conclusions are given in Sect.~\ref{SAC}. 
  

\section{Azimuth dependence of the ground contamination level}\label{ADGL}

	Stray radiation due to ground emission in the GEM experiment was initially
recognized to attain hazardous levels during test operations at the Brazilian site 
(W$44\deg 59\arcmin 55\arcsec$ -- S$22\deg 41\arcmin$) when only the rim-halo
protection had been installed. Fig.~\ref{Fig2} shows a sky map from a sample batch 
containing 123.92 hours of data taken during this testing period at \fb, where the 
horizontal striping shows clear evidence of a variable component of sidelobe 
contamination due to ground emission for the zenith-centered circularly scanning 
motion of the antenna. The map was prepared according to the same data reduction process 
that will be outlined in Sect.~\ref{DR} and included custom cuts of $60\deg$ from axis for 
the Sun, of $6\deg$ for the Moon and eventual excision of RFI signals. The 
relative calibration of the map was, however, not subjected to an adopted baseline subtraction 
technique which filters out low frequency noise. Instead, we assumed that any 
continuous set of data between successive firings of the calibrating noise source 
diode (comprising about 70\% of a full scan or, equivalently, 35\% of the  
angular extent of a great circle in the sky) would contain, at least, one pointing 
direction towards which the sky would appear uniformly cold across the entire 
declination band being mapped. This assumption is realistically incorrect, but as 
Fig.~\ref{Fig3} shows, it is nevertheless useful to portray a reasonable outline of Galactic 
features albeit an unnaturally flattened temperature distribution and some residual
stray radiation of Solar and artificial origin. The latter was absent during the test
runs only to emerge later with a 100\% duty cycle and in the direction of a near
urban area.

\begin{figure*}[hbt]
\resizebox{18cm}{!}{\includegraphics{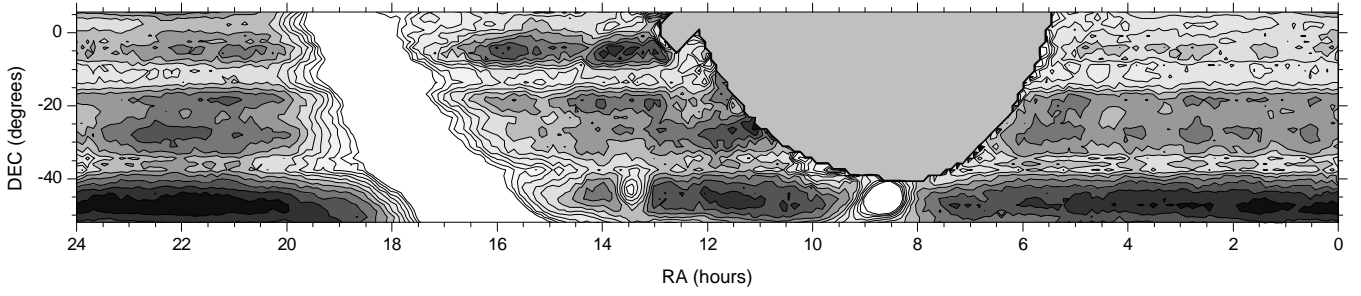}}
\caption{A sky map at \fb\ of a $60\deg$ declination band obtained with the
GEM experiment in the Southern Hemisphere using only a rim-halo protection and
assuming an uniform baseline level across the sky (Epoch $2\,000.0$ coordinates). The pixelization is $1.6\deg$
and the antenna temperature range extends 1.5\th{K} above the lowest 
temperature in 12 equally-spaced contours in order to enhance the distribution of 
Galactic radiation away from the Galactic Plane. 
	}
\label{Fig2}
\end{figure*}

\begin{figure*}[hbt]
\resizebox{18cm}{!}{\includegraphics{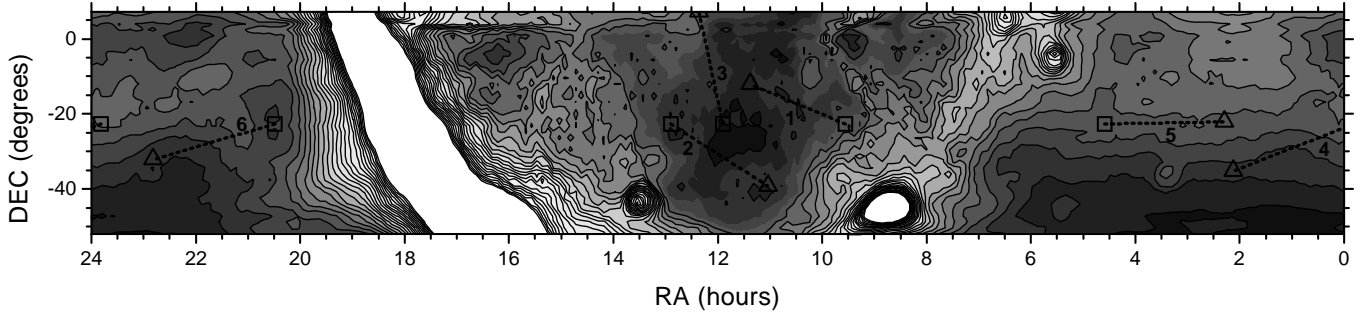}}
\caption{A GEM map at \fb\ of the same sky region as that in Fig.~\ref{Fig2} (same 
baseline assumption and pixel size), but using data obtained after the wire mesh screen 
had been added to the shielding configuration of the antenna. The antenna 
temperature ranges also over 1.5\th{K} above the lowest temperature, but the 
contour levels are spaced more tightly (60\th{mK}). Marked locations denote 
the sky directions of the 6 test measurements discussed in Sect.~\ref{TM}.
	}
\label{Fig3}
\end{figure*}

Fig.~\ref{Fig3} is a map of the same declination band as that of Fig.~\ref{Fig2} after a fence of wire 
mesh had been built around the rotating antenna. A total of 222.57 hours of data 
from an optimally-stable receiver were used in the preparation of this map. 
The azimuth-dependent contamination from the ground has been 
largely removed and we can estimate its level by subtracting 
representative azimuth scans from the two maps as described below. 
No absolute calibration of the baseline was 
attempted for either of these maps, as it is not relevant for determining differential 
measurements. This approach will enable us to refine the model used in Sect.~\ref{TM} for predicting 
a best estimate of the level of the azimuth-independent component of ground 
contamination in the survey. The locations marked in the 
map of Fig.~\ref{Fig3} correspond to the chosen 
set of sky directions, grouped pair-wise, for obtaining the differential sky 
measurements. They avoid the proximity of the Galactic Plane in order to 
diminish the chance of scale error corrections.

\begin{figure*}[hbt]
\resizebox{18cm}{!}{\includegraphics{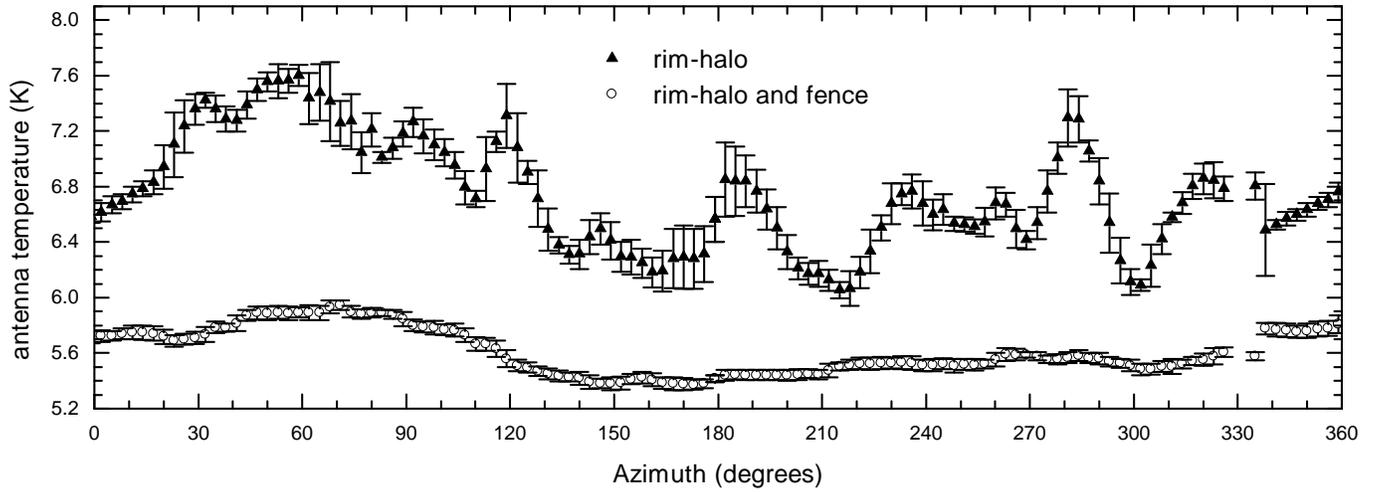}}
\caption{Antenna temperature profiles obtained before and after the construction 
of the fence and sampled along the scanning direction in regions of
relatively low sky contrast.
	}
\label{Fig4}
\end{figure*}

\begin{figure*}[hbt]
\resizebox{18cm}{!}{\includegraphics{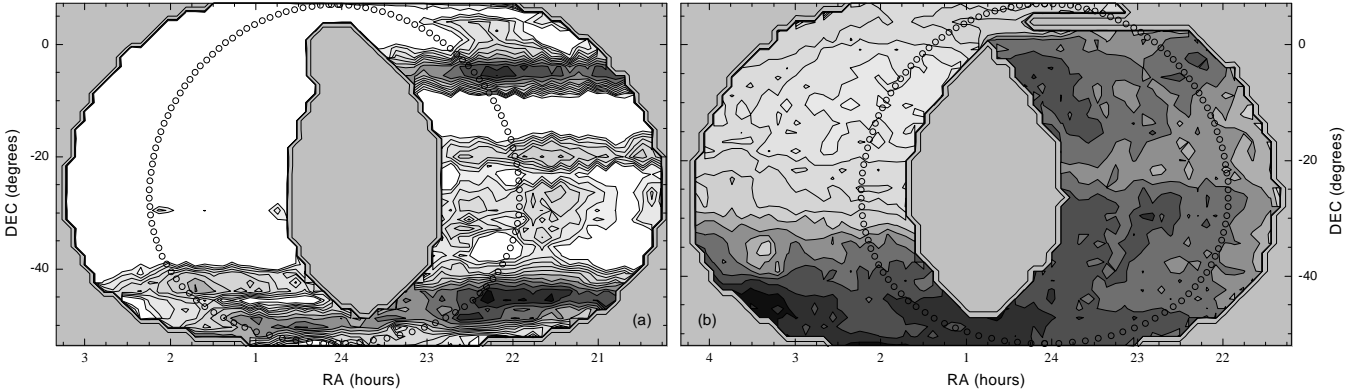}}
\caption{Maps of the sky regions revealed by the circularly scanning technique of 
the GEM experiment at \fb, (a) before and (b) after the construction of the fence,
and the distribution of the alt-azimuth sky bins used in the sampling of the 
signal displayed in the antenna temperature profiles of Figs.~\ref{Fig4}a,b. Map (a)
consumed 2.29\th{hours} of observational time and Map (b) 2.18\th{hours}. Both
maps are given at a $1.6\deg$ pixel-resolution and their antenna temperatures range
in 12 contour steps of 60\th{mK} above their respective minimum values. 
	}
\label{Fig5}
\end{figure*}

\begin{figure*}[hbt]
\resizebox{18cm}{!}{\includegraphics{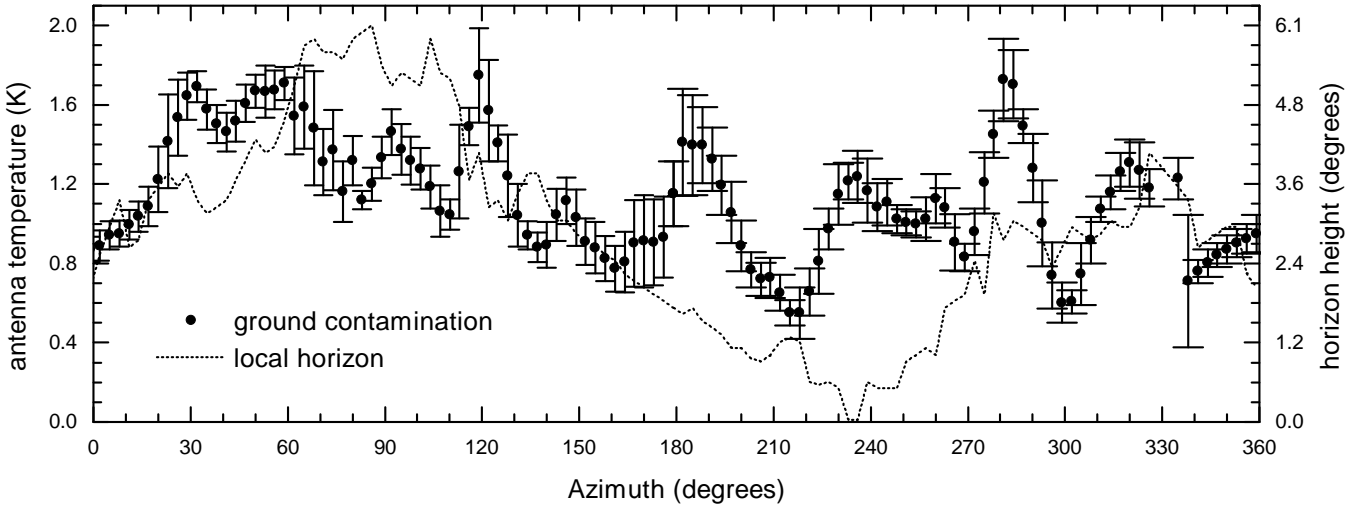}}
\caption{The level of ground contamination in the absence of the fence deduced 
from the difference in antenna temperature between the profiles of Fig.~\ref{Fig4}. The
dotted line depicts the horizon profile at the observational site.
	}
\label{Fig6}
\end{figure*}

The variable or azimuth-dependent component of ground contamination can be 
estimated by adequate
comparison of the azimuth antenna temperature profiles before and after the
introduction of the fence. Fig.~\ref{Fig4} shows two such sets of profiles. They
were obtained from single time-ordered series of scans covering the same regions 
of the sky and they sample the sky in 122 alt-azimuthal circular bins spanning 
approximately half-a-beamwidth across. This binning criteria is a basic precept 
for the relative calibration of the survey and it will not be discussed further in the 
present context. A full treatment of this calibration technique can be found in 
Tello (\cite{Te1}) and will be included in the publication of the survey. At this point we 
just mention that the series of scans were chosen for complying with highly stable 
receiver performance and relatively high Galactic latitude. This combination 
favours sky profiles of low emission contrast for easier identification of the 
ground contribution to the antenna temperature. The circular arrangement of the
sampled bins has been schematically superimposed against the observed sky in 
Figs.~\ref{Fig5}a,b. Thus the difference between the antenna temperature profiles in 
Fig.~\ref{Fig4} is a good approximation (see Fig.~\ref{Fig6}) of the ground contamination in the 
absence of the fence. It can be seen to be made up of a variable component with a 
mean amplitude of $0.52\pm0.29$\th{K} above the level of a uniform 
azimuth-independent component. The two components result from the 
convolution of the antenna beam pattern over the ground temperature 
distribution, whose spatial extent in the vertical direction is limited by the line of 
the horizon also depicted in Fig.~\ref{Fig6}. 

In the presence of the fence, we can estimate the azimuth-independent component 
of ground contamination by convolving the antenna beam pattern with an uniform 
field of radiation confined to the solid angle that the fence fills in at the prime 
focus of the parabolic dish. In this case, the beam pattern is 
the modified feed response which due to the presence of the shields gives rise to 
spillover and diffraction sidelobes. This is the subject we deal with in the next 
two sections before we assess the reality of the observations.
 
\section{Feed pattern measurements}\label{FPM}

The antenna test range of the Integration and Tests Laboratory (a satellite 
dedicated facility) at the National Institute for Space Research -- INPE -- 
was used over a period of 3 weeks in order to obtain full beam 
patterns of the GEM backfire helical feed antennas at \fa\ and \fb. For the 
measurements, a vertically polarized transmitter was located on a tower 25 m 
above the ground and 80 m in front of an anecoic chamber. The antenna under 
test sat on a plate attached to the head of the fiber glass support arm of a platform 
with 3 degrees of freedom (slide: horizontal motion along the axis between 
transmitting and receiving antennas; roll: rotation of the head support plate about 
the slide axis; azimuth: horizontal scanning motion). 

During the measurements, the upper section of the support arm was surrounded
with Eccosorb in order to avoid undesired strayed signal from the obstruction
behind the head support plate. Furthermore, since a backfire helix radiates in 
the direction of its ground plane, PVC extensions were customized to 
position the helix upside down on the head plate and to direct the feed cable 
toward its connector at the ground plane. Preliminary tests were conducted at 
different positions along the slide axis to match the phase center of the feed 
antenna with the rotation axis of the support arm. The backlobe structures of 
the feeds were also obtained by adjusting their ground planes onto the PVC 
extensions attached to the head plate.

The beam patterns were obtained by measuring the power response of the 
antennas with polar angle $\theta$ while the platform rotated through $360\deg$ 
in azimuth. The measurements were taken at $1.6\deg$ intervals at \fa\ 
and every $0.2\deg$ at \fb. The full spatial response was generated by repeating 
the azimuth scans for a sequence of equally-spaced roll angles. Although a 
$180\deg$ range in roll angle would have sufficed to cover all space directions,
the helical antenna is capable of shifting the phase of the received 
signal as it turns around its main beam axis (Kraus \cite{Kra}).
Roll angle test measurements with the \fa\ helix were consistent with this 
prediction and, in this case, the entire $360\deg$ range in roll angle was 
covered at $4.8\deg$ steps. No significant phase shifting was noticed with 
the \fb\ helix, for which $10\deg$ roll angle steps were used. 

As required by a polar angle resolution of $1\deg$ in the diffraction model we 
apply in the next section, the measured responses were regridded and interpolated 
to accomodate a $1^{\tiny\sq}$ spatial resolution. Diagrams of the resulting power 
patterns $P_{\rm n}(\theta,\phi)$ down to the 20-dB level are displayed in Figs.~\ref{Fig7}a,b.
Their mean response averaged over $\phi$ produces the pattern profiles 
$P_{\rm n}(\theta)$ shown in Fig.~\ref{Fig8}. The radiometric characterization
of these backfire helices is further illustrated in Fig.~\ref{Fig9}, showing the
antenna solid angle as a function of the polar angle $\theta$, and Table \ref{Table1} gives 
the directivity, $D$, main beam efficiency, $\epsilon_{\rm M}$, and the beam 
solid angle fraction, $\epsilon_{\rm h}$, intercepted by the co-rotating ground 
shield (halo) of the GEM parabolic reflector. The 10-dB points attain 93.8\% 
and 62.5\% of the total dish illumination at \fa\ and \fb, respectively.

\begin{figure*}[hbt]
\vspace{9.8cm}
\caption{Diagram representations of the backfire power patterns at (a) \fa\ 
and (b) \fb\ in a coordinate reference frame centered on the transmitter. 
Three concentric circles have been superimposed on the diagrams to
illustrate the opening angles of the ground shields: the boundary of the dish 
itself at $\theta=78.9\deg$, the rim-halo at $\theta=99.2\deg$ and the 
farthest lying location on the upper edge of the fence at $\theta=142.9\deg$
when $Z=45\deg$. The 10-dB level is mostly contained inside the elliptical contour. 
Arrows indicate the $\phi^\ast(\equiv 180\deg-\phi)$ orientations of the feed 
for generating the upper (upp) and lower (low) envelopes in Figs.~\ref{Fig10}--\ref{Fig13}. The subindex 
number denotes the number of shields accounted for (1 : only the rim-halo, 2 :
both halo and fence).}
\label{Fig7}
\end{figure*}

\begin{figure}[hbt]
\resizebox{8.8cm}{!}{\includegraphics{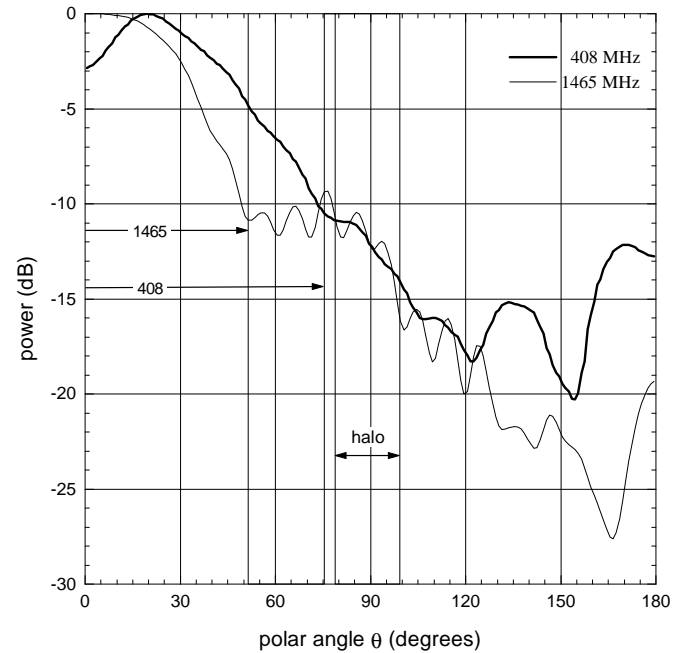}} 
\caption{Polar diagrams of the \fa\ and \fb\ feed patterns using the mean backfire 
response in the $\phi$-plane. Vertical reference lines delimit the sidelobe 
structure within the field of view of the halo and the width of the assumed
main beams.
	}
\label{Fig8}
\end{figure}

\begin{figure}[hbt]
\vspace{0.2cm}
\resizebox{8.8cm}{!}{\includegraphics{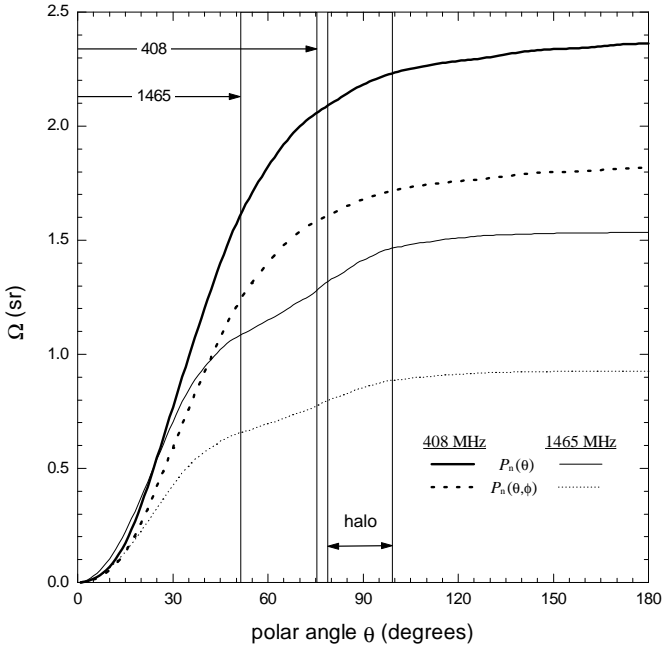}}
\caption{Antenna solid angle growth with polar angle for the beam patterns in 
Fig.~\ref{Fig8} and for the full 3-d measured response of the backfire feeds. The 
vertical lines are as in Fig.~\ref{Fig8}.
	}
\label{Fig9}
\end{figure}

\begin{table} 
\caption[]{Measured radiometric properties of backfire helical feeds}            
\begin{tabular}{cccccc}
\noalign{\vskip -6pt}                     
\hline
\noalign{\vskip 2.5pt}              
&\multicolumn{2}{c}{$P_{\rm n} (\theta)$}&&\multicolumn{2}{c}
{$P_{\rm n} (\theta,\phi)$ }\\          
\noalign{\vskip 2.5pt}         
\cline{2-3}\cline{5-6}
\noalign{\vskip 2.5pt}         
&\fa&\fb&&\fa&\fb\\
\noalign{\vskip 2.5pt}           
\hline\noalign{\vskip 4pt}           
$D$&5.32&8.19&&6.92&13.56\\
$\epsilon_{\rm M}$&0.87&0.71&&0.87&0.71\\
$\epsilon_{\rm h}\times 10^2$ &5.90&9.36&&5.89&9.32\\
\noalign{\vskip 2.5pt}           
\hline           
\noalign{\vskip 4pt}  
\end{tabular}
\label{Table1}           
\end{table}

Experimental reports on monofilar axial-mode helical antennas have seldom
focused the backfire type. End-fire helices of equivalent design characteristics,
for example, do not depend critically on frequency over the range studied here 
(see Paper I); whereas Table \ref{Table1} clearly favours a frequency dependence for the 
backfire mode.  A few authors have also attempted to describe the radiometric 
properties of the backfire helix from analytical, numerical and experimental 
points of view (Sexson \cite{Sex}; Johnson \& Cotton \cite{JC}; Nakano et al.~\cite{NYM}). 
No definite consensus has yet emerged from these studies, 
since the mechanical design of the helices under investigation was substantially 
different for each author. Our backfire feeds, which follow the design 
considerations of typical Kraus coils (Kraus \cite{Kra}), show a substantial 
narrowing of the beamwidth with increasing frequency which disagrees with 
the predictions of earlier studies (Sexson \cite{Sex}; Nakano et al.~\cite{NYM}).   


\section{Analysis of ground contamination}\label{AGC}

In the long-wavelength regime of the GEM experiment, there are two main sources of contamination, 
aside from Galactic stray radiation, which affect invariably the antenna noise temperature of the sky.
These are the emissions of the ground and the atmosphere. The latter, being a factor of at least 20\th{dB}
smaller than the former, can be safely considered to be an elevation-dependent contribution
to the signal level of the main beam. The ground contamination, on the other hand, requires a precise
knowledge of the spillover and diffraction sidelobes of the feed in order to discriminate its contribution
to the overall antenna noise temperature. In this section, we apply 
the geometric diffraction model developed in Paper I in order to account for the 
effect of shielding in the estimates of the ground signal. The shields are (see Fig.~\ref{Fig1}) a 5-m
high fence, inclined at $50\deg$ from the ground and standing at 6.4 m from the 
pivot point of the dish, and a halo of aluminum panels extending 2.1 m from the 
dish petals. The fence attenuation was estimated at the 10-dB level for \fa\ 
radiation, but below 1 dB at \fb.

\subsection{Model predictions}

Our analytical tools enable us to estimate, as a function of the zenith angle $Z$, the 
amount of ground contamination due to the unshielded and diffracted components.  
The estimates, in units of antenna temperature, are given according to Fresnel and 
Fraunhofer diffraction theories 
in order to test for near and far-field effects in the range $0\deg\le Z\le 45\deg$. 
The asymmetry of the feed patterns introduces an additional complication, since 
the solid angle over which the ground temperature is distributed (assumed to be 
the field of view below the upper edge of the fence) is seen through a sidelobe 
structure that depends on the orientation of the $\phi$-plane of the feed. 
Therefore, a family of 24 profiles was prepared for each feed by rotating the 
$\phi$-plane in $15\deg$ steps around the beam axis. For a tilted dish, the $\phi=0\deg$ reference directions 
of Fig.~\ref{Fig7} correspond to the line of sight which clears off the edge of the halo at the smallest $Z$ angle. 
Figs.~\ref{Fig10} and \ref{Fig11} display the model estimates assuming a 10 dB
attenuation from the fence (as in the 408 MHz case) in the presence and
absence of the halo, respectively.  Figs.~\ref{Fig12} and \ref{Fig13} describe the situation
of the \fb\ channel, for which the model fence provides no significant
attenuation. The upper and lower envelopes of each family of profiles have been 
identified along with some other profiles. The orientations of the \fa\ and \fb\ 
feed patterns that produce these upper and lower envelopes are indicated with 
labelled arrows in Fig.~\ref{Fig7}.

\begin{figure*}[hbt]
\resizebox{12cm}{!}{\includegraphics{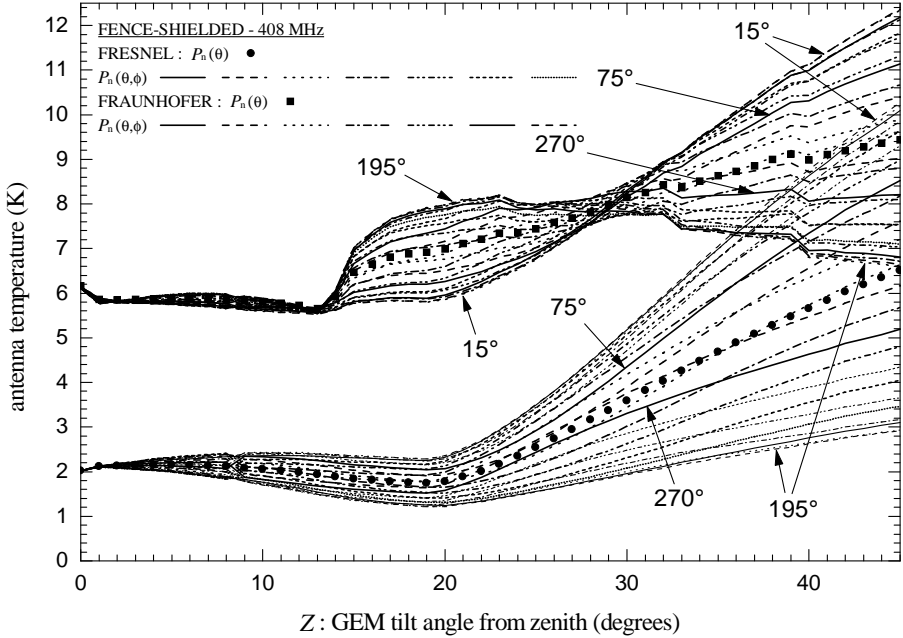}}
\hfill\parbox[b]{55mm}{
\caption{
Predicted antenna noise temperature due to transmitted and diffracted ground 
radiation in a one-shielded (fence) GEM experiment at \fa. The diffracted 
components were calculated in both the Fresnel (thin lines) and the Fraunhofer 
(thick lines) regimes. The beam pattern asymmetry of the backfire helices gives 
rise to families of profiles, some of which have been labelled according to the 
$\phi$-plane orientation of the feed. All the profiles fall into 4 sets, each of which 
has been drawn according to the sequence of line types indicated by $P_{\rm n}(\theta,\phi)$.}
\label{Fig10}
}
\vspace{36pt}
\end{figure*}

\begin{figure*}[hbt]
\resizebox{12cm}{!}{\includegraphics{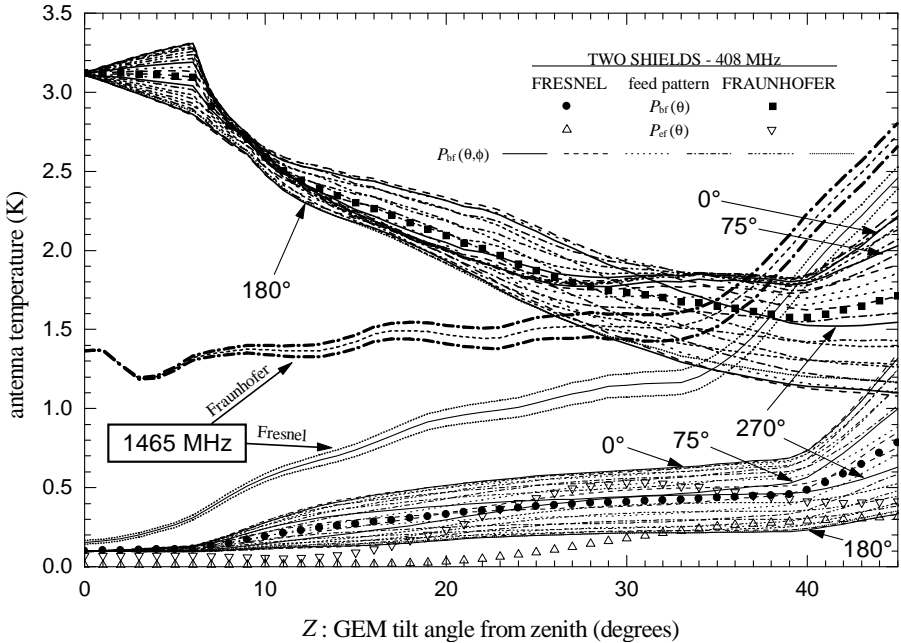}}
\hfill\parbox[b]{55mm}{
\caption{ 
Predicted antenna noise temperature due to transmitted and diffracted ground 
radiation in the double-shielded GEM experiment at \fa. Two additional triple-sets of profiles 
have been included to show the Fresnel and Fraunhofer estimates at \fb\ for $\phi=12\deg$ using a 
$7.91(^{+0.35}_{-0.32})$-dB attenuating fence and raised 80\th{cm} above the ground 
as discussed in Sect.~\ref{OTP}. Legend and label explanations are as in 
Fig.~\ref{Fig10}.}
\label{Fig11}
}
\vspace{37pt}
\end{figure*}

\begin{figure*}[hbt]
\resizebox{12cm}{!}{\includegraphics{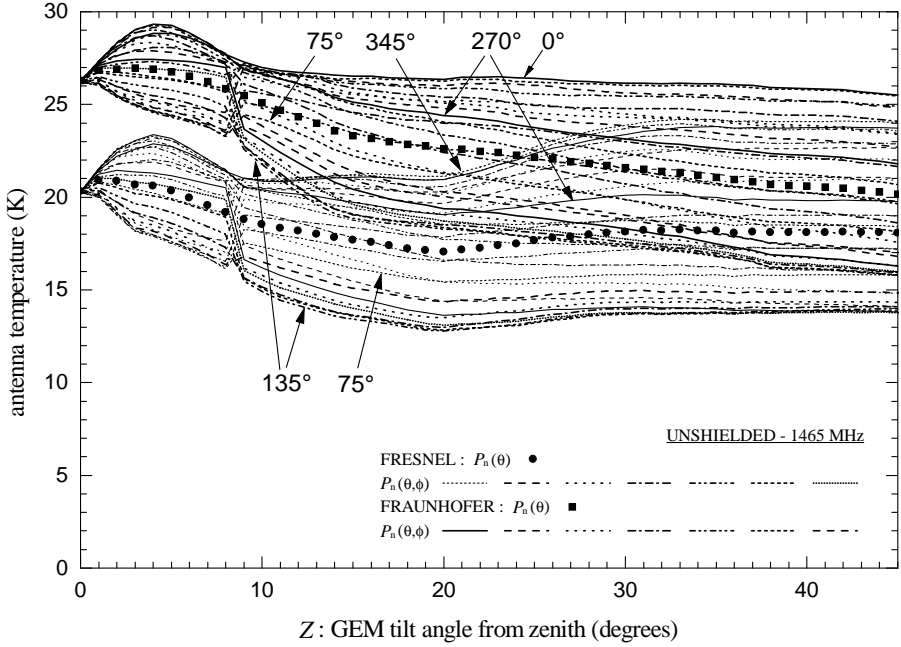}}
\hfill\parbox[b]{55mm}{
\caption{
Predicted antenna noise temperature due to transmitted and diffracted ground 
radiation in a one-shielded (no halo) GEM experiment at \fb. Legend and label 
explanations are as in Fig.~\ref{Fig10}. Since the assumed attenuation of the fence is small, 
$0.3$\th{dB}, the plotted profiles represent an effectively unshielded case.}
\label{Fig12}
}
\vspace{36pt}
\end{figure*}

\begin{figure*}[hbt]
\resizebox{12cm}{!}{\includegraphics{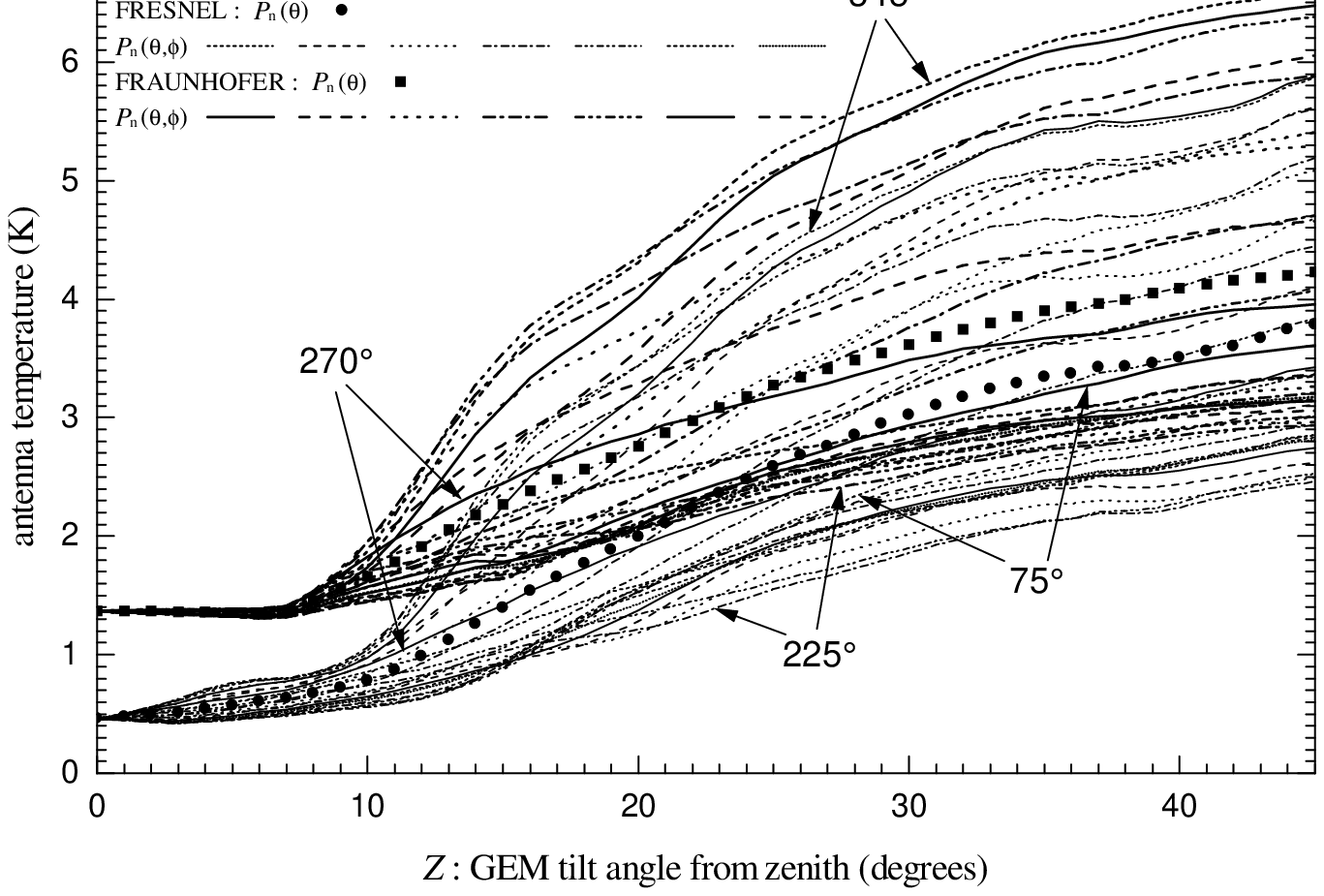}}
\hfill\parbox[b]{55mm}{
\caption{ 
Predicted antenna noise temperature due to transmitted and diffracted ground 
radiation in the double-shielded GEM experiment (an effectively one-shielded, 
fenceless, configuration) at \fb. Legend and label explanations are as in Fig.~\ref{Fig10}.}
\label{Fig13}
}
\vspace{37pt}
\end{figure*}

\subsection{Ground contamination scenarios}\label{GCS}

Figs.~\ref{Fig10} through \ref{Fig13} characterize four types of ground contamination scenarios: 
(1) fence-shielded in Fig.~\ref{Fig10}, (2) double-shielded in Fig.~\ref{Fig11}, (3) unshielded in 
Fig.~\ref{Fig12} and (4) halo-shielded in Fig.~\ref{Fig13}. The distinction is clear enough to show 
how the amount of shielding and the wavelength-dependent strength of the 
diffraction effects shape the ground contamination profiles. Thus, as we proceed 
from a weakly-diffracting and unshielded antenna scenario to a strongly-diffracting 
and double-shielded one, far-field diffraction effects give way to near-field ones. 
In doing so, the distance-dependent calculations with the Fresnel approach 
become more difficult to be matched by the Fraunhofer approximations, whose 
typical overestimating power is further increased.

Shielded scenarios produce also profiles with a tendency to resemble the 
underlying variation of the solid angle that exposes the ground for a given $Z$ 
(see Fig.~6 in Paper I). In particular, when $Z$ is large enough to expose 
unscreened ground below the fence, the corresponding profile shows a marked 
increase in ground signal. It should be noted that the profiles obtained with the 
Fraunhofer formalism in the double-shielded scenario of Fig.~\ref{Fig10} deviate from
these generalized description, since one expects the role of near-field diffraction 
at the longer wavelength and at the innermost shield to become significant.

The composition of the ground contamination profiles in terms of their transmitted 
and diffracted components can also be investigated by analyzing the symmetrized 
responses $P_{\rm n} (\theta)$. The diffraction model that 
we are using does not produce, however, separate estimates of transmitted and 
diffracted components in the Fresnel regime. Unlike in the Fraunhofer regime, 
where both components are obtained independently, the Fresnel convolution 
integral for calculating the contamination by the halo (or of the dish in the 
unshielded scenario) produces a transmission-embedded result. Nevertheless, in 
order to obtain an equivalent form of diffraction component, we have subtracted 
from the convolved result the same transmitted component as in the Fraunhofer 
regime. In a very realistic sense, the definition of a spillover sidelobe reduces to
the sidelobe level that is not modified by the presence of a physical obstruction 
along the line of sight of the feed and within the angular range of the ground 
temperature distribution. This analytical construct allows us to plot in 
Fig.~\ref{Fig14} the ratio $R_t$ of the transmitted component to the total 
ground contamination in the Fresnel regime.

The reason why the unshielded scenario in Fig.~\ref{Fig14} produces anomalous 
$R_t>1$ values is a consequence of the above given definition for the spillover 
component. This definition implies that the diffraction sidelobes (whose sidelobe 
level is modified) can actually suppress, rather than enhance, the spillover ones. 
From the point of view of a Fresnel diffraction pattern, this behavior is readily 
understood as the restriction imposed by the ground temperature distribution on 
the angular range spanning the relative power response of the feed. The 
restriction sets effectively an upper cut-off in the amplitudes of the crests that 
characterize the rippling profile of this response (see, for example, Fig.~4 in 
Paper I). Thus, if the cut-off is sufficiently low the overall relative power 
response can fall below unity. This spillover suppression is also present in the 
other scenarios, but is not dominant and, as expected from the geometrical 
argument above, it originates in the portion of the halo or dish hidden from the 
outside by the structure of the fence. The effect is stronger in the absence of the 
shields and it becomes more pronounced at the shorter wavelength. Similar
calculations with a relatively lower sidelobe structure also demonstrated that 
in order for spillover suppression to set in, the level of the relevant sidelobes
cannot be made arbitrarily small.
 
Although transmission dominates the ground contamination at large $Z$, the 
$R_t$ curves in Fig.~\ref{Fig14} indicate that diffraction becomes the dominant 
component  at lower $Z$ as the amount of shielding is also increased. We can
quantify the relevance of the spillover sidelobes by introducing a transmission 
factor $Q_{\rm n}$ (the normalized integral under the $R_t$ curves). 
Accordingly,  a thoroughly spillover-dominated scenario would result in 
$Q_{\rm n}=1$, whereas a fully diffraction-dominated case would yield 
$Q_{\rm n}=0$. Table \ref{Table2} lists the transmission factor in the four shielding
scenarios analyzed in this section. Only the double-shielded scenario may be 
recognized to be dominated by the diffraction sidelobes.

Finally, it should be stressed that the estimates given in this section have assumed
from the start that the ground temperature distribution is an isotropic field of
radiation regardless of the horizon profile. As we saw in Sect.~\ref{ADGL} this assumption 
is a valid one for a contaminating signal free of horizon-dependent variations, i.e. 
for a truly effective double-shielded scenario. Although possible, but not desirable
for experimental reasons (horizontally striped maps), the convolution of the beam
pattern with an anisotropic ground temperature distribution would yield a more
realistic estimate in the other three scenarios. In these cases, a set of profiles like
the ones shown in Figs.~\ref{Fig10}, \ref{Fig12} and \ref{Fig13} would have to be assembled for each 
particular azimuth. 

\begin{table}
\caption[]{Transmission factor $Q_{\rm n}$}            
\begin{tabular}{lccccc}
\noalign{\vskip -6pt}                     
\hline
\noalign{\vskip 2pt}                  
feed&&\multicolumn{4}{c}{Effective Shielding}\\
\noalign{\vskip 2.5pt}         
\cline{3-6}
\noalign{\vskip 2.5pt}              
regime&&none&fence&halo&double\\          
\noalign{\vskip 2.5pt}           
\hline\noalign{\vskip 4pt}           
Fresnel&&0.98&0.80&0.70&0.41\\
Fraunhofer&&0.82&0.34&0.55&0.10\\
\noalign{\vskip 2.5pt}           
\hline           
\noalign{\vskip 4pt}  
\end{tabular}
\label{Table2}           
\end{table}

\begin{figure}[hbt]
\resizebox{8.8cm}{!}{\includegraphics{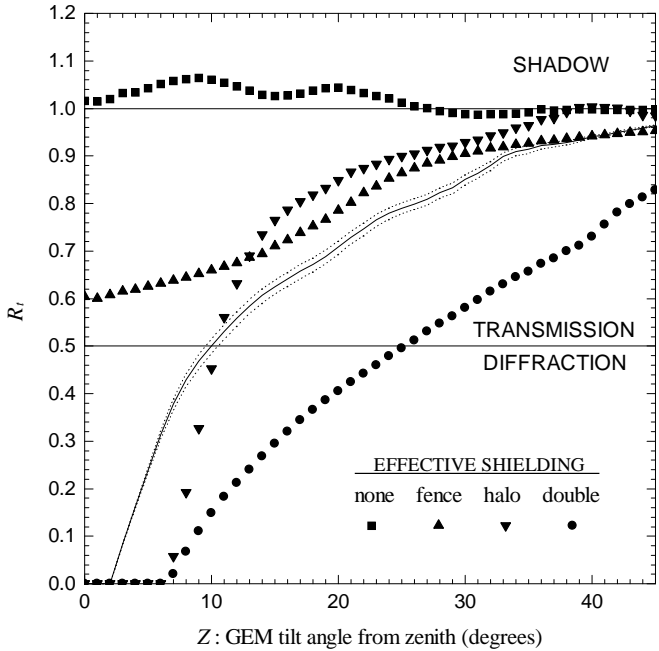}} 
\caption{The four ground contamination scenarios in terms of the ratio $R_t$ of 
the transmitted component to the total ground contamination in the Fresnel regime.
In the shadow region, spillover suppression by the diffraction sidelobes nearest to
the ground dominate the diffracted component. The line and dotted curves
mark the double-shielded case at \fb\ discussed in Sect.~\ref{FE}.}
\label{Fig14}
\end{figure}
\begin{figure*}[hbt]
\resizebox{18cm}{!}{\includegraphics{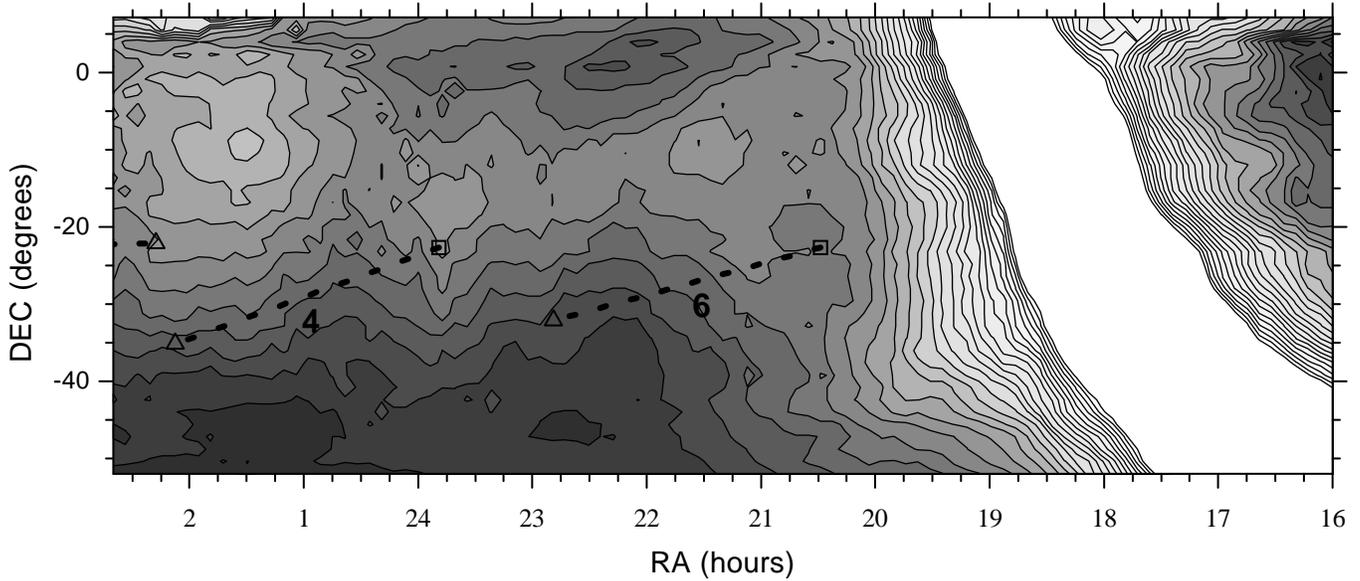}}
\caption{Destriped partial map at $1.6\deg$-pixel resolution of the  
high-sampled sky regions displayed in the declination band of Fig.~\ref{Fig3}. The 
antenna temperature range is 1.5\th{K} in 60\th{mK} contour steps. Squared 
and triangular symbols denote, respectively, the sky directions of the paired
$Z=90\deg$ and $Z=60\deg$ observations for the test measurements. 
	}
\label{Fig15}
\end{figure*}

\section{Test measurements}\label{TM}

A series of dedicated measurements was conducted with the GEM radiotelescope 
at \fb\ during the present observational period in Brazil in order to improve the
discrimination of the sky contaminating sources. The measurements consisted of 
pairs of observations taken at $Z=0\deg$ and at $Z=30\deg$ in sky directions 
away from the Galactic Plane (see Fig.~\ref{Fig3}). Each observation sampled the 
radiometric signal 
every 0.56 seconds over a few minutes while an approximate 15-minute interval 
elapsed between the $Z=0\deg$ and the $Z=30\deg$ samplings. In this manner, a 
total of 6 measurements were obtained over a nearly 3-month period. Although 
the absolute level of ground contamination in general will be somewhat different 
for different pairs, the mean difference between the two levels, 
$\bar\Delta T_{\rm A}$, can be used for comparison with the model predictions 
outlined in the preceding section.  

This differential measurement approach relies, however, on our ability to separate 
likewise the other constituents of the antenna noise temperature, namely, the 
atmospheric emission and the sky background. The latter is a mixture of synchrotron
and free-free radiation originating in the Galaxy, Cosmic Microwave Background 
Radiation (CMBR) and a diffuse background of extragalactic origin. Depending on 
the sky direction Galactic emission at \fb\ can be some 5 times larger or even a 
full order of magnitude smaller than the signal due to the CMBR. The atmospheric 
contribution, on the other hand, is necessarily larger at $Z=30\deg$ than at the 
zenith because of a larger air mass. At \fb\ the bulk of the emission by the 
atmosphere is due to the pressure-broadened spectra of the O$_2$ molecule. 
Using the reference model proposed by Danese and Partridge (\cite{DP}) 
(see also Liebe \cite{Lie} and Staggs et al.~\cite{Sta}) a straightforward secant 
law correction to the zenith contribution at the Brazilian site gives an estimate for 
the differential atmospheric component of $0.305\pm0.090$\th{K}.

\subsection{Data reduction}\label{DR}

Our data was first time-ordered and corrected for thermal susceptabilities of the 
receiver baseline ($0.3591\pm0.0007$\th{K/$\deg$C}) and fractional gain
($0.00922\pm0.00001/\deg$C). Then, 44.8-second bursts of 2.24-second firings 
of a thermally stable noise source diode were extracted from the data stream and 
used to calibrate the overall system gain. Table \ref{Table3} summarizes the results of the 
observations along with the number of samples and the implied differences in 
antenna temperature between the two $Z$ directions for: ({\it i}) the 
measurements, ({\it ii}) the Galactic emission background and ({\it iii}) the final 
budget (including the increase due to the larger optical depth of the atmosphere at 
$Z=30\deg$). The Galactic contribution was estimated using a partial map (65.21
hours of data) of the sky signal from the GEM experiment at \fb, whose baseline
has been so far properly corrected according to a destriping algorithm in 
order to filter out low frequency noise (Tello \cite{Te1}). The data for this map
makes up about 30\% of the data used in preparing the map in Fig.~\ref{Fig3}, but due to
sampling differences (which bias the destriping process -- see also Table \ref{Table4}) it has 
been split into the two maps shown in Figs.~\ref{Fig15} and \ref{Fig16} along with the locations 
chosen for the paired measurements listed in Table \ref{Table3}.

\begin{table*}
\caption[]{Antenna temperature in the GEM experiment at \fb\ 
for observations at $Z=0\deg$ and $30\deg$.}            
\begin{tabular}{ccrrcrrcccc}
\noalign{\vskip -6pt}                     
\hline
\noalign{\vskip 2pt}                  
&&\multicolumn{2}{c}{$Z=0\deg$}&&\multicolumn{2}{c}{$Z=30\deg$}
&&\multicolumn{3}{c}{$\Delta T_{\rm A}$\th{(K)}}\\
\noalign{\vskip 2.5pt}         
\cline{3-4}\cline{6-7}\cline{9-11}
\noalign{\vskip 2.5pt}              
pair&&\multicolumn{1}{c}{$T_{\rm A}\pm 1\sigma$\th{(K)}}&\multicolumn{1}{c}{$N$}
&&\multicolumn{1}{c}{$T_{\rm A}\pm 1\sigma$\th{(K)}}&\multicolumn{1}{c}{$N$}
&& measurement & Galaxy & final budget\\
\noalign{\vskip 2.5pt}           
\hline\noalign{\vskip 4pt}           
1 && $10.798\pm0.042$ & 104 && $12.151\pm0.035$ & 150 && $1.353\pm0.055$
   & $-0.083\pm0.035$ & $0.97\pm0.11$\\
2 && $9.620\pm0.036$ & 146 && $11.028\pm0.036$ & 93 && $1.408\pm0.051$
   & $-0.254\pm0.066$ & $0.85\pm0.12$\\
3 && $11.212\pm0.040$ & 219 && $12.180\pm0.046$ & 347 && $0.967\pm0.061$
   & $+0.307\pm0.043$ & $0.97\pm0.12$\\
4 && $7.923\pm0.035$ & 71 && $9.299\pm0.040$ & 148 && $1.376\pm0.054$
   & $+0.173\pm0.050$ & $1.24\pm0.12$\\
5 && $8.922\pm0.030$ & 132 && $10.151\pm0.031$ & 82 && $1.229\pm0.043$
   & $-0.028\pm0.011$ & $0.90\pm0.10$\\
6 && $12.472\pm0.038$ & 131 && $13.582\pm0.039$ & 267 && $1.110\pm0.054$
   & $+0.185\pm0.024$ & $0.99\pm0.11$\\
\noalign{\vskip 2.5pt}           
\hline           
\noalign{\vskip 4pt}  
\end{tabular}           
\label{Table3}
\end{table*} 

\begin{figure}[hbt]
\resizebox{8.8cm}{!}{\includegraphics{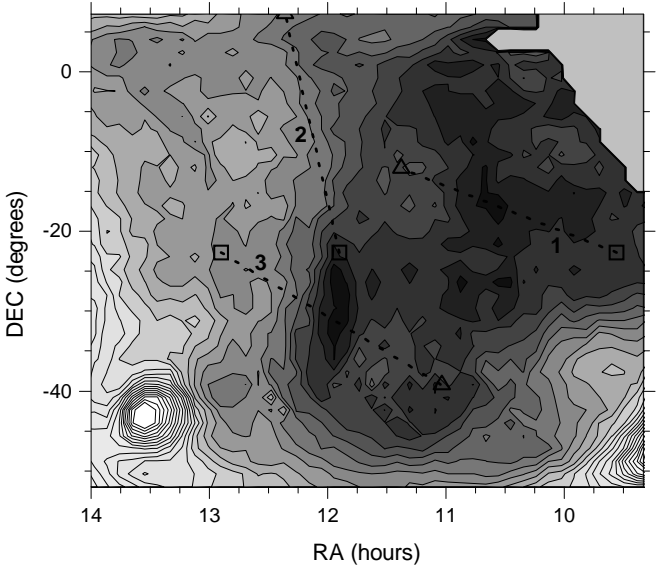}} 
\caption{The low-sampled complement of the map in Fig.~\ref{Fig15}, but at the
same resolution and with the same gray scaling in antenna temperature. The
upper-right hand corner is data defficient due to $60\deg$ custom cuts around
the Sun.}
\label{Fig16}
\end{figure}

In order to extract the antenna temperature
in a given direction, the pixel nearest to it was found first and then averaged with
the surrounding set of 8 neighbouring pixels taken at half-weights. This procedure
allows us to sample the sky in a square region $4.8\deg$ ($1.6\deg$ per pixel) on the side and 
is consistent with a HPBW of $\approx 5.4\deg$ for the \fb\ beam (Tello \cite{Te1}).
This can also be verified in Table \ref{Table4} where we compare these estimates with those 
of the nearest pixel value itself and the average from the 4-pixel area enclosing the 
given direction along with the sampling differences among the different pairs. Note
that pair 5 is actually missing in Fig.~\ref{Fig15} and, therefore, we have provisionally 
supplemented the data in Tables 3 and 4 with the differential measurement obtained
using the map in Fig.~\ref{Fig3}. To see that this is not as bad as it appears,  the mean 
absolute difference between the estimates for pairs 1, 2 and 3 in the maps of Figs.~\ref{Fig3}
and \ref{Fig16} (low-sampled sky) is $0.237\pm0.066$\th{K}, but only $0.112\pm0.040$\th{K} 
for pairs 4 and 6 in the high-sampled regions of the map in Fig.~\ref{Fig15}. Thus, 
within the sensitivity of our measurements ($\approx 20$\th{mK}) the Galactic 
contributions to the differential measurements in Table \ref{Table3} turn out to be smaller 
than, or as large as, the one estimated for the emission of the atmosphere.

\begin{table*}
\caption[]{Effects of binning strategy for the Galactic contribution to differential 
measurements at \fb $^{\mathrm{a}}$ }
\begin{tabular}{cccrrcrrrcrrr}
\noalign{\vskip -6pt}                     
\hline
\noalign{\vskip 2pt}                  
&&\multicolumn{3}{c}{1-pixel}&&\multicolumn{3}{c}{4-pixel matrix}
&&\multicolumn{3}{c}{9-pixel matrix}\\
\noalign{\vskip 2.5pt}         
\cline{3-5}\cline{7-9}\cline{11-13}
\noalign{\vskip 2.5pt}              
pair&&\multicolumn{1}{c}{$\Delta T_{\rm A}$\th{(K)}}&
\multicolumn{1}{c}{$N_{90}$}&\multicolumn{1}{c}{$N_{60}$}
&&\multicolumn{1}{c}{$\Delta T_{\rm A}\pm 1\sigma$\th{(K)}}&
\multicolumn{1}{c}{$N_{90}$}&\multicolumn{1}{c}{$N_{60}$}
&&\multicolumn{1}{c}{$\Delta T_{\rm A}\pm 1\sigma$\th{(K)}}&
\multicolumn{1}{r}{$N_{90}$}&\multicolumn{1}{r}{$N_{60}$}\\
\noalign{\vskip 2.5pt}           
\hline\noalign{\vskip 4pt}           
1 && $-0.100$ & 13 & 9 && $-0.078\pm0.032$ & 48 & 39 &&
	$-0.083\pm0.035$ & 110 & 92 \\ 
2 && $-0.288$ & 4 & 63 && $-0.257\pm0.026$ & 14 & 233 &&
	$-0.254\pm0.066$ & 33 & 338 \\
3 && $+0.254$ & 18 & 11 && $+0.325\pm0.034$ & 74 & 56 &&
	$+0.307\pm0.043$ & 165 & 120 \\
4 && $+0.222$ & 70 & 47 && $+0.217\pm0.023$ & 293 & 203 &&
	$+0.173\pm0.050$ & 697 & 428 \\
5 && $-0.008$ & 160 & 252 && $-0.019\pm0.006$ & 637 & 997 &&
	$-0.028\pm0.011$ & 1419 & 2179 \\
6 && $+0.178$ & 63 & 85 && $+0.177\pm0.023$ & 251 & 343 &&
	$+0.185\pm0.024$ & 568 & 784 \\
\noalign{\vskip 2.5pt}         
\hline\noalign{\vskip 4pt}
\multicolumn{2}{c}{$\bar\Delta T_{\rm A,\oplus}^{\rm obs}$} &
\multicolumn{3}{c}{$1.003\pm0.071^{\mathrm{b}}$ }&&
\multicolumn{3}{c}{$0.983\pm0.046\pm0.054$}&&
\multicolumn{3}{c}{$0.992\pm0.044\pm0.062$}\\
\noalign{\vskip 2.5pt}           
\hline           
\noalign{\vskip 2.5pt}  
\end{tabular}
\begin{list}{}{}
\item[$^{\mathrm{a}}$] The entries referred to as $N_{90}$ and $N_{60}$
correspond to the number of observations sampled in the 
\item[] determination of the sky signal observed toward $Z=90\deg$ and 
$Z=60\deg$, respectively.
\item[$^{\mathrm{b}}$] Only the external error ($\sigma/\sqrt{5}$) has been assigned
in this case.
\end{list}
\label{Table4}           
\end{table*}

The weighted average of the values in the last column of Table \ref{Table3} is an estimate of 
the differential ground contamination in the GEM experiment at \fb. We obtain 
$\bar\Delta T_{\rm A,\oplus}^{\rm obs} = 0.992$\th{K} with internal and external 
1-$\sigma$ error estimates of 0.044 and 0.062\th{K}, respectively (see also Table \ref{Table4}). 
Based on the ratio between these two errors, we can rule out the presence of 
systematic errors, which may have been introduced, for instance, by unaccounted
stray radiation contamination of sidelobes other than those considered here. In fact,
aside from the differential measurement approach, which reduces the effect of
of residual sidelobe contamination, the signal contrast of even the brightest sky 
features relative to that of the ground does not go above the 13-dB level. Only the
presence of the Sun could offer potential problems, but except for pair 1, none
of the other measurements was conducted with the Sun above the horizon. Still,
the estimate from pair 1 does not raise suspicious concerns, eventhough the Sun
was seen at $90.0\deg$ from axis and at $71.2\deg$ during the observations 
toward $Z=60\deg$ and $Z=90\deg$, respectively. 

\subsection{Orientation of the $\phi$-plane}\label{OTP}

Before attempting a comparison of $\bar\Delta T_{\rm A,\oplus}^{\rm obs}$ 
with our model predictions, we need
to assign the orientation of the $\phi$-plane of the feed in order to select
the most likely profile. In addition, we have to apply the model calculations
for the shield configuration actually used during the observations. Although the
halo was the same as the one assumed to obtain the results in Figs.~\ref{Fig10} through
\ref{Fig13}, the attenuation of the fence was increased by using a wire mesh with holes half
as small and wires 25\% thinner (according to our attenuation formula in Paper I
we should thereby obtain a 6.2-dB screening effect at \fb). Finally, the entire 
fence was raised 80\th{cm} above the ground.

The orientation of the $\phi$-plane of the feed could be inferred by direct 
comparison of the feed diagram in Fig.~\ref{Fig7}b with the mapping of the beam pattern 
of the antenna by some convenient point source. This procedure is, of course, 
based on the assumption that the feed axis is also not perfectly aligned with the
optical axis of the secondary for an asymmetric beam pattern to be projected
onto the sky.  In our case we chose the Sun, at a particular time of the year, 
which at the Brazilian site can be made to intercept the Galactic scans at 
$Z=30\deg$ with sufficient angular coverage ($\sim 30\deg$) around the beam 
axis. The result of such a mapping is displayed in Fig.~\ref{Fig17} in 20 contour steps of 
1\th{dB}. The brightest region, corresponding to the precise passage of the scan circle 
through the Sun, could not be mapped up to a true 0-dB level because the signal 
overshot the detector threshold. This may have caused the double-lobed structure 
seen inside the main beam pattern in Fig.~\ref{Fig7}b to smooth out in the mapping of 
Fig.~\ref{Fig17}. In fact, in 1994, when the solar activity was relatively low, we recorded a
solar transit (see Fig.~\ref{Fig18}) in Bishop, CA, which did not saturate the detector and 
did reveal a double-peaked main beam. In Fig.~\ref{Fig17} the innermost contours follow 
the outlines of a bulged shape which is reminiscent of the double-lobed structure. 
Thus, together with the ellipticity of the surrounding contours in both diagrams 
we determined the $\phi$-plane orientation of the feed from the difference in 
the orientation of the major axis of these elliptical contours. The 10-dB contours 
are well confined inside elliptical boundaries with eccentricities of 0.64 and 0.34 
for the feed and antenna patterns, respectively. The semi-major axis of the ellipse 
in the direction of the larger lobe in Fig.~\ref{Fig7}b is then oriented along $\phi=125\deg$ 
while that in the direction of the bulged region in Fig.~\ref{Fig17} corresponds to 
$\phi^\ast=317\deg$. Since $\phi^\ast\equiv\phi+180\deg$, according to the system 
of coordinates used in Fig.~\ref{Fig17}, we obtain a $\phi$-plane orientation for the feed of 
$12\deg$.

\begin{figure}[hbt]
\resizebox{8.8cm}{!}{\includegraphics{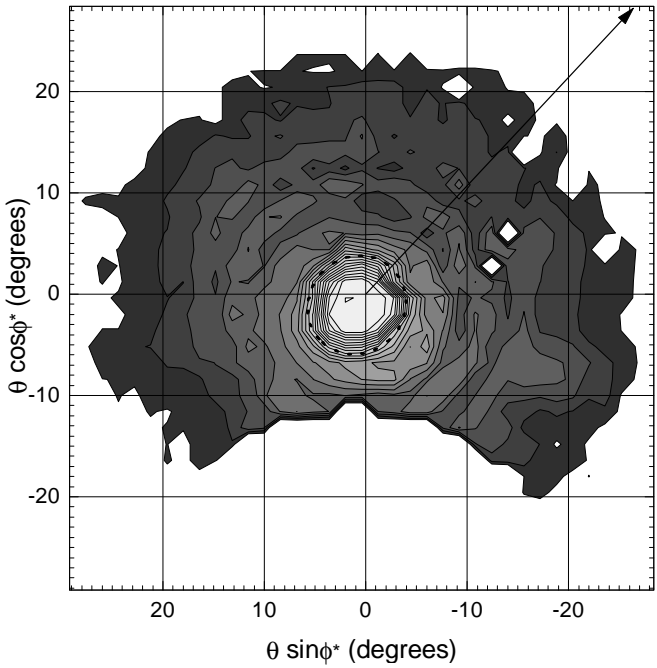}} 
\caption{Beam pattern mapping of the \fb\ backfire-fed GEM antenna, in twenty 1-dB steps
and at a pixel resolution of at $1.6\deg$, using the passage of the Sun through the $Z=30\deg$ scan 
circles of the antenna on the 29-th of September 1999 in Cachoeira Paulista, Brazil. 
The $\phi^\ast$ angle of the pattern is measured counterclockwise from the 
ordinate axis and corresponds to $\phi-180\deg$ in the
coordinate system of Fig.~\ref{Fig7}b while the elevation of the Sun is given by 
$60\deg-\theta\cos\phi^\ast$. The arrow indicates the major axis alignment of the 
10-dB elliptical contour toward the larger component of the double-lobed 
structure in Fig.~\ref{Fig7}b.
	}
\label{Fig17}
\end{figure}

\begin{figure}[hbt]
\resizebox{8.8cm}{!}{\includegraphics{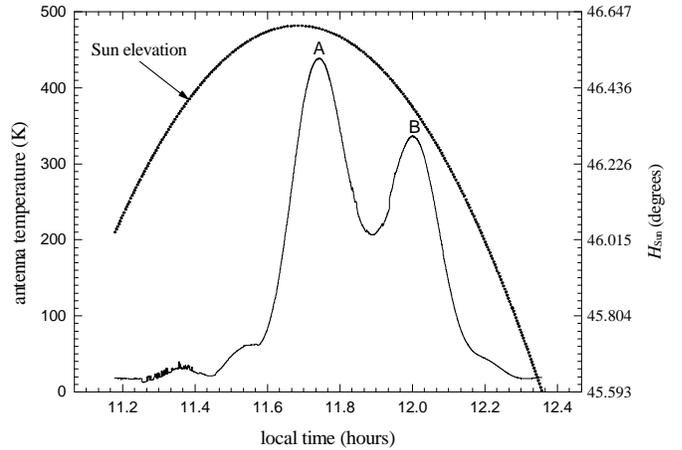}}
\caption{Antenna noise temperature record (solid line) of a solar transit in Bishop, 
CA, during data  taking operations at \fb\ in October 1994. The elevation of the 
Sun, $H_{\rm Sun}$, is given by the dotted line. The peaks A  and  B indicate 
relative maxima in  the antenna response and span an $\approx 5\deg$ interval in 
the azimuth coordinate of the Sun.
	}
\label{Fig18}
\end{figure}

\subsection{Final estimates}\label{FE}

Our diffraction model predicts a differential ground contamination of $\Delta 
T_{\rm A,\oplus} = 1.380$\th{K} for the shield configuration used during the 
observations and an orientation of $\phi_{\rm plane}=12\deg$. In order to 
predict the observed value of $\bar\Delta T_{\rm A,\oplus}^{\rm obs}$, 
we have to adjust the attenuation coefficient of the wire mesh by an efficiency factor 
$\beta=0.675\pm0.052$ or, equivalently, increase the screening of the fence by 
$1.71^{+0.35}_{-0.32}$\th{dB}. The resultant profile has been included in Fig.~\ref{Fig11}.
$\beta$ scales linearly not only with $\Delta T_{\rm A,\oplus}$, but also with the 
predicted differential ground contributions from the halo, 
$\Delta T_{\rm A,\oplus}^{\rm\,hal}$, and from the fence,
$\Delta T_{\rm A,\oplus}^{\rm\,fen}$. So, if
\beqy\label{beta}
	{\beta\over10^{-3}} = -155.880 + 837.363 
\,\left({\Delta T_{\rm A,\oplus}\over {\rm K}}\right)\;,
\eeqy
then the corresponding contributions from the halo and from the fence are
\beqy\label{hal}
	\left({\Delta T_{\rm A,\oplus}^{\rm\,hal}\over {\rm mK}}\right)
 = 31.49 - 169.25\,\left({\Delta T_{\rm A,\oplus}\over {\rm K}}\right)
\eeqy
and
\beqy\label{fen}
	\left({\Delta T_{\rm A,\oplus}^{\rm\,fen}\over {\rm mK}}\right)
= 193.61 - 40.00\,\left({\Delta T_{\rm A,\oplus}\over {\rm K}}\right)
\eeqy
with 1-$\sigma$ errors of 0.03 and 0.01 in the zero-points and linear 
coefficient, respectively, in (\ref{hal}) and (\ref{fen}); but 1 order of magnitude
smaller in those of (\ref{beta}).

These formulae tell us that, as the screening of the fence becomes less
efficient ($\beta$ increasing), the differential ground contribution increases,
eventhough the one from the diffracted components decreases. In this 
spillover-dominated scenario with $Q_{\rm n}=0.67\pm0.01$ (see
Fig.~\ref{Fig14}) the ground contamination contributed by diffraction at the halo and
at the fence will decrease with increasing $Z$ as long as $\beta\gsim0.00011$ 
and $\beta\gsim3.9$, respectively. For most practical fences, the lower bound on 
$\beta$ implies that diffraction at the halo should always decrease with $Z$. In 
order to have the same scenario at the fence, the attenuation of the wire mesh 
would have to be quite low ($\lsim0.3$\th{dB}).

Table \ref{Table5} gives the refined model estimates of the ground contamination levels
for GEM observations at \fb\ in the Southern Hemisphere.\footnote{The 1-st
part of an all-sky GEM survey at \fb\ is presently in preparation and combines
Northern Hemisphere observations with Southern data to cover nearly 75\%
of the sky. Both data sets were obtained using $Z=30\deg$ scans only.}

\begin{table}
\caption[]{Ground contamination in $Z=30\deg$ GEM data at \fb\ from the 
Southern Hemisphere.}            
\begin{tabular}{cllcccc}
\noalign{\vskip -6pt}                     
\hline
\noalign{\vskip 2pt}                  
&\multicolumn{1}{c}{sidelobe}&\multicolumn{1}{c}{shield$^{\mathrm{a}}$}
&&\multicolumn{1}{c}{contamination}&&\multicolumn{1}{c}{error}\\
&&&&\multicolumn{1}{c}{(mK)}&&\multicolumn{1}{c}{(mK)}\\
\noalign{\vskip 2.5pt}         
\hline\noalign{\vskip 4pt}           
&spillover &double&& 975 && 75 \\
&diffraction & fence&& 154 && \multicolumn{1}{l}{\hspace{2.5ex}3} \\
&diffraction & halo I&& \multicolumn{1}{l}{\hspace{6.9ex}28} 
&& \multicolumn{1}{l}{\hspace{2.5ex}2} \\
&diffraction & halo II&& \multicolumn{1}{l}{\hspace{6.1ex}-11} 
&& \multicolumn{1}{l}{\hspace{2.5ex}1} \\
\noalign{\vskip 2.5pt}         
\hline\noalign{\vskip 4pt}
&Total &double&& 1146 && 75 \\
\noalign{\vskip 2.5pt}           
\hline           
\end{tabular}
\begin{list}{}{}
\item[$^{\mathrm{a}}$] Estimates are given for a double-shielded scenario
were the rim-halo contribution to the diffracted component has been separated
into exposed (halo I) and hidden (halo II) portions as discussed in 
Sect.~\ref{GCS}. 
\end{list}
\label{Table5}                      
\end{table}

\section{Summary and conclusions}\label{SAC}

Levelling and reducing the contamination of the antenna temperature by ground 
emission is an important requirement in survey experiments for mapping the 
non-thermal component of the Galactic emission background. In the 
zenith-centered 1-rpm circular scans of the GEM experiment this is achieved by 
using a wire mesh fence around a rim-halo shielded antenna. 
Without the fence, a prohibitive variable component 
of ground contamination compromises the data taken with this portable 5.5-m
dish in the Southern Hemisphere at \fb\ with a mean amplitude of 
$0.52\pm0.29$\th{K} above the level of a uniform azimuth-independent 
component. With the fence, the level of a uniform component was obtained 
by comparing differential measurements of the antenna temperature 
toward selected regions of the sky with model predictions of the spillover and 
diffraction sidelobes. 

First of all, the model allowed us to investigate the shielding performance of the 
experiment using the fully measured beam patterns of the GEM backfire helical 
feeds at \fa\ and \fb. We concluded that far-field diffraction effects dominate a 
weakly-diffracting and unshielded antenna scenario whereas near-field 
effects dominate a stronger-diffracting and double-shielded scenario. 
Furthermore, the shielding efficiency of the experiment could be quantified in terms 
of the normalized cumulative ratio $Q_{n}$ of the spillover-induced transmission 
to the overall sidelobe contamination in the zenith angle range 
$0\deg\le Z\le 45\deg$. If the shielding is low enough, spillover sidelobe 
suppression will ensue, since the ground temperature angular distribution can 
introduce an upper cut-off in the relative power response of the feed. A critical
element in the analysis is introduced, however, by the need to account for the 
assymetric response of the feed and which seems, most likely, to result from 
imperfect alignment of the feed axis on the measuring stand and along the optical 
axis of the secondary. We used the near sidelobe pattern (out to some $30\deg$ 
from axis) of the radiotelescope to ressolve the issue. 

Finally, we applied atmospheric and Galactic corrections to 
the differential measurements before comparing the residual signal with the
model predictions for the level of ground contamination. The choice of sky
directions away from the Galactic Plane led to contributions from the sky between
$Z=0\deg$ and $Z=30\deg$ which were as high, but not larger, than the ones 
expected from the emission of the atmosphere. The former were derived from a 
template sky with a sensitivity of 20\th{mK} based on GEM data taken at \fb\ in 
the Southern sky with a HPBW$\approx 5.4\deg$.

The corrected test measurements match the model predictions if we introduce 
a screening efficiency factor $\beta$ which shows strict and separate linear 
correlations with the differential ground contamination and its diffraction 
components generated at the shields. 
Consequently, it suffices that the (total) differential ground contamination 
be known, for its spillover and diffracted components to be identified uniquely.
With the refined model ($\beta=0.675\pm0.052$) a uniform level of ground 
contamination is estimated at $1.146\pm0.075$\th{K} with a spillover-to-diffraction 
component ratio of $5.7\pm0.5$. This is a spillover dominated scenario with 
$Q_{\rm n}=0.67\pm0.01$ and decreasing diffraction sidelobes with increasing 
$Z$.

\begin{acknowledgements}
      The authors are particularly in debt to Alexandre M.~Alves, Luis Arantes, 
Edson R.~Rodrigues, Agenor P.~da Silva and Rog\'erio R.~de Souza for technical and 
observational support. We are also grateful to the LIT-INPE Antennas Group 
for its collaboration during the feed pattern measurements. 
The GEM project in Brazil is presently being supported by FAPESP through
grants 97/03861-2 and 97/06794-4. M.~Bersanelli acknowledges the support of the
NATO Collaborative Grant CRG960175. 
\end{acknowledgements}


\begin{thebibliography}{}

\bibitem[1991]{Ban} Banday A.J., Wolfendale A.W., 1991,
MNRAS 248, 705

\bibitem[1992]{Be1} Bennett C.L., Smoot G.F., Hinshaw G., et al., 1992, 
ApJ 396, L7

\bibitem[1996]{Be2} Bennett C.L., Smoot G.F., Hinshaw G., et al., 1996, 
ApJ 464, L1

\bibitem[1972]{Bhj} Berkhuijsen E.M., 1972 , 
A\&AS 5, 263

\bibitem[1989]{DP} Danese L., Partridge R.B., 1989, 
ApJ 342, 604

\bibitem[1996a]{Dav} Davies R.D., Gutierrez C.M., Hopkins J., et al., 1996a,
MNRAS 278, 883

\bibitem[1996b]{DWG} Davies R.D., Watson R.A., Guti\'errez C.M., 1996b,
MNRAS 278, 925

\bibitem[1994]{Ami} De Amici G., Torres S., Bensadoun M., et al., 1994, 
ApSS 214, 151

\bibitem[1998] {Del} Delabrouille J., 1998, 
A\&AS 127, 555

\bibitem[1995]{Gun} Gundersen J.O., Lim M., Staren J., et al., 1995, 
ApJ 443, L57

\bibitem[1996] {Har} Hartmann D., Kalberla P.M.W., Burton W.B., Mebold U., 
1996 A\&AS 119, 115

\bibitem[1997] {HB} Hartmann D., Burton W.B., 1997 Atlas of Galactic Neutral
Hydrogen. Cambridge University Press.

\bibitem[1970]{Ha1} Haslam C.G.T., Quigley M.J.S., Salter C.J., 1970, 
MNRAS 147, 405    
   
\bibitem[1974]{Ha2} Haslam C.G.T., Wilson W.E., Graham D.A., Hunt G.C.,
1974, A\&AS 13, 359
   
\bibitem[1981]{Ha3} Haslam C.G.T., Klein U., Salter C.J., et al., 1981, 
A\&AS 100, 209 

\bibitem[1984]{JC} Johnson R.C., Cotton R.B., 1984, IEEE Trans.~Antennas 
Propagat.~32, 1126 

\bibitem[1999]{Jon} Jones A.W.: 1999, Foregrounds and experiments below
33 GHz. In: de Oliveira-Costa A., Tegmark M.~(eds.) Microwave Foregrounds. 
PASPC 181, 321 


\bibitem[1996a]{Ko1} Kogut A., Banday A.J., Bennett C.L., et al., 1996a, 
ApJ 460, 1

\bibitem[1996b]{Ko2} Kogut A., Banday A.J., Bennett C.L., et al., 1996b,  
ApJ 464, L5

\bibitem[1988]{Kra} Kraus J.D., 1988, Antennas, McGraw-Hill, New York, 
Chap.~7

\bibitem[1987]{Law} Lawson K. D., Mayer C. J., Osborne J. L., Parkinson M.L., 
19987, MNRAS 225, 307 

\bibitem[1985]{Lie} Liebe H.J., 1985, Rad.~Sci.~20(5), 1069

\bibitem[1996]{Lim} Lim M., Clapp A.C., Devlin M.J., et al., 1996, 
ApJ 469, L69

\bibitem[1999] {Lop} L\'opez-Corredoira M., 1999, 
A\&A 346, 369  

\bibitem[1999]{Mai} Maino D., Burigana C., Gorski K., et al., 1999, 
A\&AS 140, 383

\bibitem[1988]{NYM} Nakano H., Yamauchi J., Mimaki H., 1988,
IEEE Trans.~Antennas Propagat.~36(10), 1359

\bibitem[1998]{Pla} Platania P., Bensadoun M., Bersanelli M., et al., 1998, 
ApJ 505, 473

\bibitem[1982]{Rei} Reich W., 1982,  
A\&AS 48, 219

\bibitem[1986]{RR} Reich P., Reich W., 1986, 
A\&AS 63, 205

\bibitem[1965]{Sex} Sexon T.L., 1965, Sylvannia Electron.~Syst.~Rep.~ASTIA 
Doc.~626088

\bibitem[1992]{Sm1} Smoot G.F., Bennett C.L., Kogut A., et al., 1992,
ApJ 396, L1

\bibitem[1999]{Sm2} Smoot G.F.: 1999, CMB synchrotron foreground. In: de
Oliveira-Costa A., Tegmark M.~(eds.) Microwave Foregrounds. 
PASPC 181, 61

\bibitem[1996]{Sta} Staggs S.T., Jarosik N.C., Wilkinson D.T., Wollack E.J., 1996,
ApJ 246, 407

\bibitem[1997]{Te1} Tello C., 1997, Um experimento para medir o brilho total do c\'eu
em comprimentos de onda centim\'etricos, Tese de doutorado, Instituto Nacional
de Pesquisas Espaciais, S\~ao Jos\'e dos Campos, SP, Brasil

\bibitem[1999]{Te2} Tello C., Villela T., Wuensche C.A., et al., 1999, 
Rad.~Sci.~34(3), 575

\bibitem[1996]{Tor} Torres S., Ca\~non V., Casas R., et al., 1996, 
ApSS 240, 225

\end{thebibliography}
\end{document}